\documentclass{informat} 
\usepackage{infothm}
\usepackage{times}
\usepackage[T1]{fontenc}
\usepackage{multirow}
\usepackage{listings}
\usepackage{color}
\usepackage{bold-extra}
\usepackage{amssymb}
\usepackage{soul}
\usepackage{graphicx}

\definecolor{Brown}{cmyk}{0,0.81,1,0.60}
\definecolor{OliveGreen}{cmyk}{0.64,0,0.95,0.40}
\definecolor{CadetBlue}{cmyk}{0.62,0.57,0.23,0}
\definecolor{MyDarkBlue}{rgb}{0,0.08,0.45}

\newcommand{\comment}[1]{}

\newcommand{\lstref}[1]{Listing~\ref{#1}}
\newcommand{\figref}[1]{Figure~\ref{#1}}
\newcommand{\secref}[1]{Section~\ref{#1}}

\newcommand{\tool}[1]{\textsc{#1}}
\newcommand{\ATF}{\textsc{Grammatic}$^{PG}$}

\lstdefinelanguage{Grammatic}
	{
		morestring=[b]',
		morekeywords={lex,empty,*,?,+,before,after,at},
		morecomment=[l]{//},
	}
\lstdefinelanguage{Typesystem}
	{
		morestring=[b]',
		morekeywords={typesystem,language,for,backend,type},
		morecomment=[l]{//},
	}
\lstdefinelanguage{ANTLR}{
	morecomment=[s][\itshape\color{MyDarkBlue}]{\{}{\}},
	morekeywords={returns,int},
	morestring=[b]',
	morecomment=[l][\color{red}]{//!},
}

\begin{document}
\volume{}
\issue{}
\pubyear{2010}
\firstpage{}
\lastpage{}
\articletype{}   
\begin{frontmatter}                           
%
\title{EXTENSIBLE TYPE CHECKER\\ FOR PARSER GENERATION}
\runningtitle{Extensible type checker for parser generation}
\author{\fnms{Andrey} \snm{BRESLAV}}
\address{Mathematics Department at Natural Science Faculty,\\
St. Petersburg State University of Information Technology, Mechanics and Optics\\
Kronverskiy~49, 195009, St.Petersburg, Russia\\
\email{abreslav@gmail.com}
}
\runningauthor{A. Breslav}
%
\begin{abstract}
	\emph{Parser generators} generate translators from language specifications. In many cases, such specifications contain \emph{semantic actions} written in the same language as the generated code. Since these actions are subject to little static checking, they are usually a source of errors which are discovered only when generated code is compiled.

	In this paper we propose a parser generator front-end which statically checks semantic actions for typing errors and prevents such errors from appearing in generated code. The type checking procedure is extensible to support many implementation languages. An extension for Java is presented along with an extension for declarative type system descriptions.
\end{abstract}
\end{frontmatter}


\maketitle

\section{Introduction.}\label{Introduction}

\begin{figure}[h!]
		\centering
		\includegraphics[width=.8\textwidth]{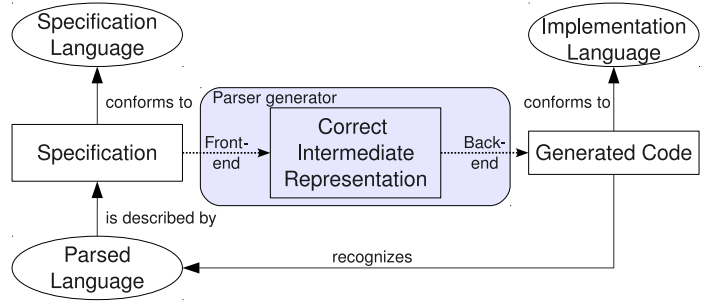}
		\caption{Generator structure and languages involved}\label{languages}
\end{figure}

Parser generators have been used for decades to create translators and other language processing tools.
A parser generator is a tool that reads a specification of a language and generates code which is capable of checking its input text for syntactical correctness and construct an internal representation from it. This process involves several languages (see \figref{languages}): the specification is written in a \emph{specification language} and describes the \emph{parsed language}, while the generated code (which recognizes the parsed language) is written in an \emph{implementation language} (e.g., Java or C). 

A specification language usually includes a notation for context-free grammars and some means to specify \emph{semantic actions} which are the computations translating a program into the internal representation. In this paper we consider \emph{on-line} parser generators such as \tool{Yacc} \cite{YACC}, \tool{ANTLR} \cite{ANTLR}, \tool{Coco/R} \cite{Coco/R} and \tool{JavaCC} \cite{JavaCC}, which are characterized by having the semantic actions defined on the concrete grammar, as opposed to attribute grammar (AG) systems such as \tool{Eli} \cite{Eli} and \tool{JastAdd} \cite{JastAdd} which define the computations on abstract syntax trees (ASTs) and usually require the complete input to be parsed before the computations are started. 

A parser generator translates a grammar specification into a parser and integrates user-defined semantic actions into it.
Since the user may make mistakes while writing the actions, this frequently leads to generating code that contains errors which are reported only by the implementation language compiler. We will now illustrate this problem with an example of a simple language of arithmetic expressions with variables. The \tool{ANTLR} grammar for this language is shown in \lstref{arithexp} (no semantic actions are presented at this point). 
\begin{lstlisting}[label=arithexp,caption=An ANTLR grammar for arithmetic expressions]
// Lexical rules
fragment LETTER : 'a'..'z' | 'A'..'Z' | '_' ;
fragment DIGIT  : '0'..'9' ;
VAR    : LETTER (LETTER | DIGIT)* ;
INT    : DIGIT+ ;
// Syntactic rules
expr   : term (('+' | '-') term)* ;
term   : factor ('*' factor)* ;
factor : VAR | INT | '(' expr ')' ;
\end{lstlisting}

Let us consider the case when one uses \tool{ANTLR} to develop a parser that checks the syntax and evaluates the expressions in a given environment. Evaluation of an expression is a special case of translation: the expression in essentially translated into a number. An environment stores values for all variables referenced in the expression. For example, if the parser is run on the input text ``\texttt{x*(3+2)}'' in the environment [x=4], it accepts the input and returns 20. When run on ``\texttt{(x+*3)}'', it does not accept (raises an exception), because the input is not syntactically correct. 
Note that the notation in \lstref{arithexp} does not describe environments: an environment is passed as a separate argument to the parser (see the examples below). 

To give an example of an error which may appear in the generated code, let us consider the following set of semantic actions for the rule \texttt{factor}:
\begin{lstlisting}[language=ANTLR]
factor[Environment env] returns [int result]
	: VAR { result = env.getValue($VAR.getText()); }
	| INT { result = $INT; }
	| '(' e=expr[env] ')' { result = e; } ;
\end{lstlisting}
When we run \tool{ANTLR} on a specification containing this rule it will successfully generate code. The code generated for the second alternative will contain the following lines:
\begin{lstlisting}[language=Java,escapechar={!}]
	int result = 0;
	// ...
	Token INT2=null;
	// ...
	result = INT2;
\end{lstlisting}
The Java compiler yields an error message at the last line: a value of type \texttt{Token} can not be assigned to a variable of type \texttt{int}. Now we have to figure out that the cause of this error is that we forgot to extract the contents of the token by calling the \texttt{getText()} method on \texttt{\$INT} in the specification. Let us correct this error:
\begin{lstlisting}[language=ANTLR]
	| INT { result = $INT.getText(); }
\end{lstlisting}
We run \tool{ANTLR} again and in the generated code the erroneous line changes to the following:
\begin{lstlisting}[language=Java]
	result = INT2.getText();
\end{lstlisting}
The Java compiler complains again at the same line: a \texttt{String} can not be assigned to an \texttt{int} variable. We have to correct the specification again:
\begin{lstlisting}[language=ANTLR]
	| INT { result = Integer.parseInt($INT.getText()); }
\end{lstlisting}
After generating the code again, we can see that it compiles successfully. 

This process took us three complete runs of \tool{ANTLR}, each followed by a compilation attempt, and we had to analyze the generated code twice to figure out which part of the specification causes the error. The development process has a form of the cycle shown in \figref{cycles} (left side). The generated code is usually hard to read and, in case of long specifications, code generation may take up to several seconds. Thus, this cycle may be time-consuming. 

\begin{figure}[h!]
\centering
\framebox{
\begin{minipage}{.45\textwidth}
\begin{enumerate}
\setlength{\itemsep}{0pt}
		\item Change the specification.
		\item Generate the code.
		\item Try to compile the code.
		\item Get error messages in terms of the implementation language.
		\item Read the generated code to trace the errors back to their causes in the specification.
		\item Go to 1.
\end{enumerate}
\end{minipage}
}
\framebox{
\begin{minipage}{.45\textwidth}
\begin{enumerate}
\setlength{\itemsep}{0pt}
		\item Change the specification.
		\item Try to generate code.
		\item Ge error messages in terms of the specification language.
		\item Go to 1.
\end{enumerate}
\vspace{52pt}
\end{minipage}
}
\caption{Development cycles: conventional and with static checking}\label{cycles}
\end{figure}

The main motivation of the present study is to reduce the development cycle to the one given in \figref{cycles} (right side).
It corresponds to the usual way of working with compilers: change-compile-errors-change. This paper aims at achieving this goal by improving static checking of parser specifications.

A parser generator may be logically divided into two parts (see \figref{languages}): a \emph{front-end} which reads a specification, performs the static checks and builds an internal representation of a parser, and a \emph{back-end} which generates code. To support the shortened development cycle, only the front-end may be allowed to report errors to the user. The back-end must silently produce code which must be error-free as soon as the front-end did not find any errors in the specification. 

Most generators have front-ends which check only for errors which prevent them from building the internal representation, such as usage of undefined names. This leads to the problem described above.

In this paper we present an extensible specification language and a front-end infrastructure which detects type incompatibilities in assignments and function calls. In combination with name-screening and import statement generation in the back-end, this prevents all types of compiler errors for implementation languages like Java, C\#, C or Pascal. 

Our approach is designed under the requirement of being applicable for many implementation languages, and thus the generic specification language may be extended to support a particular type system. We present an implementation of such a type system for Java and a generic language which allows one to specify simple type systems declaratively.

We report on \ATF{} --- a prototype implementation of our approach built on top of \tool{ANTLR}. \ATF{} specifications are type-checked and transformed into \tool{ANTLR} specifications. The \tool{ANTLR} tool-chain can then be used to generate the parser. 

The rest of the paper is organized as follows. \secref{Overview} gives an overview of our specification language illustrating the language design and basic constructs. Extensible type system of \ATF{} and a demonstration of how the development cycle changes with it is presented in \secref{TypeSystem}. This type system is parameterized by the types of an implementation language and corresponding subtyping rules, and thus it must be \emph{instantiated} to be actually used. In \secref{DefaultExtension} we present a mechanism for declaratively describing type systems of implementation languages and instantiating \ATF{} using them. This mechanism describes the type systems as having a fixed number of types with subtyping rules specified explicitly. To provide a better integration with a particular implementation language, one can write a custom extension. The extension mechanism and an example extension supporting Java are described in \secref{Java}. As we mentioned before, the front end provides an internal representation of a parser from which a back-end can generate error-free code. We show how the latter is done in our prototype in \secref{Backend}. \secref{RelatedWork} describes related work, and \secref{Conclusion} summarizes our contribution and points out directions for future work.


\section{Overview of the specification language.}\label{Overview}

Attribute grammars (AGs) introduced in \cite{ATG} are very convenient for describing translators, but in a general case they require the whole input to be parsed before the translation starts. This is why the most popular parser generators only use some concepts of AGs, but not the whole framework.

\ATF{} specification language corresponds to a restricted version of AGs where each nonterminal symbol $N$ is associated with a set of attributes, which is divided into
\begin{itemize}
	\item \emph{output} attributes: computed by the productions defining $N$;
	\item \emph{input} attributes: computed by the productions which use $N$.
\end{itemize}
Output attributes of \ATF{} directly correspond to \emph{synthesized} attributes used in AGs, and input attributes represent a restricted case of \emph{inherited} attributes. 

\subsection{Translation functions and external functions.}

We use the terms ``input'' and ``output'' for attributes because they correspond very closely to inputs and outputs of functions. One can think of a \emph{syntactic rule} defining a nonterminal (which is a set of all productions for this nonterminal) as a ``\emph{translation function}''. The productions which use the nonterminal \emph{call} this function, passing the input attributes to is (as arguments). The function itself computes the output attributes and \emph{returns} them to be used by the caller. A recursive descent parser implements this analogy in a one-to-one manner: for each nonterminal there is a corresponding function which takes the input attributes as parameters and returns the output attributes. 

In principle, there may be many translation functions corresponding to the same syntactic rule. In \ATF{} notation, one specifies these functions right after the rule. A translation function is described by its \emph{signature} (a name and lists of input and output attributes) and a body (this will be explained later), for example:
\begin{lstlisting}
	N : ... ; // Syntactic rule
		translateN(int in) --> (int out) { // Signature
			// Body
		}
\end{lstlisting}
Note that all attributes are declared with their types. 

Being a recursive descent parser generator, \tool{ANTLR} also uses a function analogy for translation, although it does not support many translation functions for the same rule. In \tool{ANTLR} notation the example given above looks as follows:
\begin{lstlisting}[language=ANTLR]
	N[int in] returns [int out] : ... ;
\end{lstlisting}

Unlike general AGs (and like \tool{ANTLR}), \ATF{} prescribes the order of computations using translation schemes \cite{DragonBook}. In other words, attributes are computed by \emph{semantic actions} positioned somewhere inside productions. The most popular notation for translation schemes (it is used by the majority if not all parser generators, including \tool{ANTLR}) is the following:

$A \rightarrow B \,\{action\} \,C$

This notation is rather intuitive but in practice (e.g., in large \tool{ANTLR} grammars) it makes the specifications unreadable because of mixing context-free productions and action code. To avoid this problem, in \ATF{} notation we separate the grammar productions from the actions by using a technique similar to \tool{AspectJ}'s \emph{advice} \cite{AspectJ}. The above mentioned production in our notation can be written as follows:

\begin{lstlisting}[emph={action},emphstyle={\it}]
	A : B C;
		translateA(...) --> (...) {
			...
			after B : action ;
			...
		}
\end{lstlisting}

The action is specified in the body of a translation function. The \textbf{after} keyword denotes that the action must be executed after the nonterminal $B$ is matched. Another option to specify the same behaviour is to use the \textbf{before} keyword as follows:

\begin{lstlisting}[emph={action},emphstyle={\it}]
	before C : action ;
\end{lstlisting}

The actions to the right of the ``:'' sign can assign values to attributes and call \emph{external functions} to perform computations. External functions must be implemented outside the specification, in \ATF{} they are represented only by their signatures, for example
\begin{lstlisting}
	add(int x, int y) --> (int sum);
\end{lstlisting}
A developer must supply the implementation written separately (in the implementation language).

In addition to specifying which actions to execute before or after a certain position, one needs to say which translation function to call for a nonterminal on the right-hand side of the syntactic rule, what arguments to pass to it and where to store the returned values. For this purpose we use the \textbf{at} keyword followed by a name of the nonterminal:
\begin{lstlisting}
	at C : tmp = translateC(a, b) ;
\end{lstlisting}
This, written inside a translation function for \texttt{A}, means that for the occurrence of \texttt{C} on the right-hand side of the syntactic rule for \texttt{A} we must call the translation function \texttt{translateC}, passing arguments \texttt{a} and \texttt{b} and writing the returned value to \texttt{tmp}.

To illustrate some technical details of the \ATF{} notation, we will now proceed to the example of arithmetic expressions mentioned in the \secref{Introduction}.

\subsection{Example.}

The grammar for arithmetic expression is given in \lstref{arithexp}. Our translation functions must evaluate an expression in a given environment which contains values for the variables used in the expression. To do this, we will need the appropriate external functions:
\begin{lstlisting}
	strToInt(String s) --> (Int value);
	value(Environment env, String variable) --> (Int value);
	zero() --> (Int zero);
	one() --> (Int one);
	neg(Int x) --> (Int negx);
	add(Int x, Int y) --> (Int sum);
	mul(Int x, Int y) --> (Int prod);
\end{lstlisting}
We do not fix an implementation language here: the example described in this section will work for any implementation language in which types can have names, and \texttt{Int}, \texttt{String} and \texttt{Environment} can be defined with the straightforward meanings.

We are going to specify three translation functions: one for each of the nonterminals \texttt{factor}, \texttt{term} and \texttt{expr}. Each of these functions must have an input attribute for the environment and an output attribute for the result. The translation function for \texttt{factor} is shown in \lstref{factorTF}.
\begin{lstlisting}[
	label=factorTF,
	caption=Translation function for \texttt{factor},float=htbp]
factor : VAR | INT | '(' expr ')' ;           
  factor(Environment env) --> (Int result) { 
    after VAR : result = value(env, VAR#); // External function 
    after INT : result = strToInt(INT#);   // External function
    at expr   : result = expr(env);        // Translation function
  }
\end{lstlisting}

The actions for \texttt{VAR} and \texttt{INT} use the notation ``\emph{NAME}\#'' which denotes a textual value of a token (this value is of type \texttt{String}). Note that \texttt{value} and \texttt{strToInt} which are called \emph{after} the tokens, are external functions, while \texttt{expr} is a translation function called \emph{at} the occurrence of the corresponding nonterminal. One can not do anything but a call to a translation function in \textbf{at} position. In some cases calls to translation functions (and thus, whole \textbf{at}-actions) may be omitted, but in our example this is not the case: we have to specify what argument to pass to the translation function \texttt{expr}.

The rule for \texttt{term} is different from what we have shown by now in the sense that it has two occurrences of the same nonterminal \texttt{factor}. At both these occurrences we need to call the translation function \texttt{factor} defined above and pass the environment object to it. The result will be assigned to a \emph{local attribute} \texttt{f}, which is used inside the translation function as an auxiliary storage. This attribute \texttt{f} is declared inside the translation function to have the type \texttt{Int}:
\begin{lstlisting}
	Int f;
	at factor: f = factor(env);
\end{lstlisting}
This action is common for both occurrences of factor, but \emph{after} the value is acquired we must treat it differently in each case. To be able to distinguish between different occurrences of the same nonterminal, one can use \emph{location labels} available in \ATF{}. With the use of such labels, the translation function for factor looks as follows:
\begin{lstlisting}[
	label=termTF,
	caption=Translation function for \texttt{term},float=htbp]
term : $f1=factor ('*' $f2=factor)* ;         
	term(Environment env) --> (Int result) {  
		Int f; // Local attribute declaration
		at factor : f = factor(env);      // For both occurrences
		after $f1 : result = f;              // Only for $f1
		after $f2 : result = mul(result, f); // Only for $f2
	}
\end{lstlisting}

A label may be attached not only to a nonterminal occurrence, but also to any phrase in a production. This feature is rather convenient when defining a translation function for \texttt{expr} (see \lstref{exprTF}). This function also give an example of a \emph{block} which groups several statements together (see the action for \texttt{\$t1}).

\begin{lstlisting}[
	label=exprTF,
	caption=Translation function for \texttt{expr},float=htbp]
expr : $t1=term ( $sgn=('+' | '-') term)* ;         
	expr(Environment env) --> (Int result) {  
  		at     term : t = term(env);
		before $t1  : {
			result = zero();
			sign = one();
		}
		after  term : result = add(result, mul(sign, t));
		before $sgn : sign = one();
		after  '-'  : sign = neg(sign);
  }
\end{lstlisting}

To motivate again our choice of notation, we provide the same rule in \tool{ANTLR} notation (in \lstref{exprANTLR}). As can be seen, it is rather hard to understand the structure of the syntactic rule from it, which is never the case for \ATF{}.
\begin{lstlisting}[
	label=exprANTLR,language=ANTLR,
	caption=\tool{ANTLR} rule for \texttt{expr},float=hb]
expr[Environment env] returns [int result] 
	: {result = 0; sign = 1;} 
	  t=term[env] {result = t;}
	  (
 	    {sign = 1;} ('+' | '-' {sign = -1;}) 
	    t=term[env] {result += t * sign;}
	  )* 
	;         
\end{lstlisting}

\subsection{Multi-return functions.}
All functions used in our example have only one output attribute, but in general a function may return a tuple. For example, a function \texttt{divide(x,~y)} may return two numbers: a quotient and a remainder. \ATF{} does not support tuple-typed attributes, and a return value of this function can not be assigned to a single attribute. Instead, \ATF{} supports attribute tuples. If one needs to receive a result of the \texttt{divide} function, it can be done in the following way:
\begin{lstlisting}
Int quot;
Int rem;
(quot, rem) = divide(x, y);
\end{lstlisting}
This code assigns the first component of a returned tuple to the attribute \texttt{quot} and the second --- to the attribute \texttt{rem}.

\subsection{Attribute initialization.}

Before the first assignment, a value of an attribute is not defined and thus can not be read. To ensure that every attribute is initialized before the first usage, \ATF{} performs conventional data-flow analysis \cite{DataFlow}.
If the analysis finds a read-access which may not be preceded by a corresponding write-access, the front-end reports an error. This analysis relies on the construction of a control flow graph. \ATF{} does not support conditional operators and loops as such, and all the branching and repetition happens according to the structure of grammar rules which is denoted by the common regular operations: concatenation (sequence), alternative (``$|$'') and iteration (``+''). Optional constructs (``?'' and ``*'') are viewed as alternatives with an empty option.

\begin{figure}[h!]
	\includegraphics[width=\textwidth]{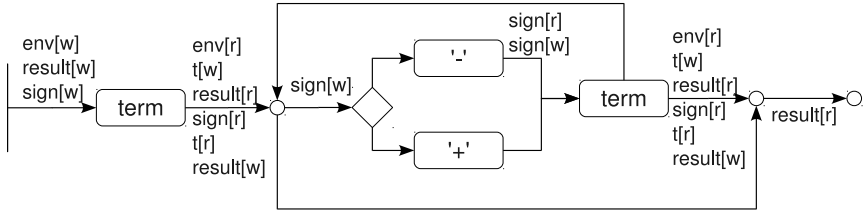}
	\caption{Control flow graph for \texttt{expr} with attribute access labels}\label{cfg}
\end{figure}

A control flow graph is constructed as follows: concatenation corresponds to sequential execution, alternative corresponds to branching and iteration corresponds to a loop. \figref{cfg} shows a control flow graph for the \lstref{exprTF}; edges are labeled with corresponding sequences of attribute reads and writes indicated as ``[r]'' and ``[w]'' respectively.

\subsection{Local type inference.}\label{TypeInf}

\ATF{} checks translation functions for type safety: when a value is assigned to an attribute (passing as an argument can also be interpreted as an assignment to an input attribute of a function), the type of the right-hand side must be a subtype of the type of the left-hand side. This prevents \ATF{} from generating code with typing errors such as those we discussed in \secref{Introduction}.

In \lstref{exprTF}, two attributes, \texttt{env} and \texttt{result}, are declared in the signature of the translation function to have types \texttt{Environment} and \texttt{Int} respectively, but the intermediate attribute \texttt{t} is not declared anywhere. What type does it have? This is an example of \emph{local type inference} which makes \ATF{} feel more like dynamic languages: if some attribute is used but not declared, the type checker assumes that it is a local attribute and tries to figure out an appropriate type for it considering the context in which it is used. 

In our example,  \texttt{t} is first assigned the value returned by the \texttt{term} function, which has an output attribute \texttt{result} of type \texttt{Int}. We write this as follows:
\begin{equation}\label{teqres}
		t \leftarrow result^{\mathtt{Int}}
\end{equation}
Then, it is passed to the \texttt{mul} function as a value for an input parameter \texttt{y} of type \texttt{Int}:
\begin{equation}\label{yeqt}
		y^{\mathtt{Int}} \leftarrow t
\end{equation}
These two usages facilitate the following reasoning: assuming that there exists a type $\tau$ for \texttt{t}, such that the whole translation function is typed correctly, from~(\ref{teqres}) we see that \texttt{Int} must be a subtype of $\tau$, and from (\ref{yeqt})  we see that $\tau$ must be a subtype of \texttt{Int}. Hence, $\tau$ equals \texttt{Int}, and we have inferred the type for \texttt{t} successfully. 

The type checker in \ATF{} applies this kind of reasoning for every attribute which is used but not declared. In some cases this procedure does not lead to a definitive conclusion. In these cases the type checker reports an error, which can be reconciled by providing an explicit declaration of the attribute in question. 

Not only attribute types but also signatures of external functions can be inferred in this manner. For example, assume the following assignment appears in a specification:
\begin{lstlisting}
a = f(b, c)
\end{lstlisting}
If \texttt{f} is not declared, \ATF{} assumes that there is an external function \texttt{f} with one output attribute and two input attributes, and applies the above reasoning to these attributes. If it succeeds, a complete type for \texttt{f} is inferred.

We provide a more detailed description of type checking and type inference in \ATF{} in the next section.


\section{Type system.}\label{TypeSystem}

This section describes the extensible type system used in \ATF{}. The typing rules presented below are written under the assumption that all the attributes and external functions are declared explicitly. The purpose of type inference in this case is to reconstruct omitted annotations. If the reconstruction is not possible, the specification is considered to be inconsistent.

\subsection{Typing rules.}

\newcommand{\G}{ {\mathcal{G}_L} }
\renewcommand{\L}{L}
\newcommand{\eql}{\cong_\L}
\newcommand{\lel}{\le_\L}
\newcommand{\gel}{\ge_\L}
\newcommand{\TUP}{TupleType}
\newcommand{\Str}{String_\L}

Let the implementation language be denoted by $\L$. The types of the implementation language will be denoted $\G$ and referred to as \emph{ground types}. Let $Str \in \G$ be the ground type which represents character strings. The types used in \ATF{} are defined by the following productions:
$$
\begin{array}{rcl}
	AttributeType &::=& \G \\
	TupleType &::=& (AttributeType^*) \\
	FunctionType &::=& TupleType \rightarrow TupleType\\
\end{array}
$$
As the names suggest, attributes may only have ground types, tuples are sequences of attributes and have corresponding types, and functions send tuples to tuples. As we explained above, attribute tuples are used to receive return values of functions having more than one output attribute. Note that tuples can not be nested. Function arguments and individual return attributes can only have ground types: for example, a single argument can not be a tuple.

A \emph{subtyping relation} on the set $\G$ of ground types (denoted $\lel$) represents subtyping rules of the implementation language. We assume that it is reflexive and transitive. We will use $\tau \gel \sigma$ and $\sigma \lel \tau$ interchangeably.

Each translation function is type-checked separately. A type-checking context $\Gamma$ comprises signatures of all functions available in the specification along with declarations of all input and output attributes of these functions and local attributes of the function being checked. Since attributes in different signatures may have same names, each attribute is indexed with the name of the function it belongs to. For example, a context may contain the following declarations:\\
$$\begin{array}{rcl}
factor&:& (Environment) \rightarrow (Int)\\
env^{factor} &:& Environment\\
result^{factor} &:& Int\\
term &:& (Environment) \rightarrow (Int)\\
env^{term} &:& Environment\\
result^{term} &:& Int\\
\end{array}$$

\figref{exptypes} provides straightforward typing rules for token values, attributes and tuples and a rule for function application which says that a type of an argument must be a subtype of the type of the corresponding formal parameter.

\newcommand{\trule}[3]{%
\dfrac{#1}{#2}(\mbox{\textsc{#3}})
}

\begin{figure}[htbp]
$$\begin{array}{lll}

	\begin{array}{l}
		\trule{}{\Gamma \vdash \mathtt{NAME\#} : \Str}{token}\\
		\\
		\trule{}{\Gamma, \alpha : \tau \vdash \alpha : \tau }{attribute}\\
	\end{array}
&&
\trule{
\begin{array}{l}
	\tau_1 \ldots\tau_n \in \G\\
	\Gamma \vdash x_1 : \tau_1 \, \ldots \, \Gamma \vdash x_n : \tau_n
\end{array}
}{
	\Gamma \vdash (x_1, \ldots, x_n) : (\tau_1, \ldots, \tau_n)
}{tuple}

\\
&&\hspace{10pt}\\
\multicolumn{3}{c}{
\trule{	
\begin{array}{lll}
		\tau_1 \ldots\tau_n,\, \sigma_1 \ldots \sigma_n,\, \rho_1 \ldots \rho_m \in \G
		&\quad&
		\sigma_1 \gel \rho_1\,\ldots\, \sigma_m \gel \rho_m\\
		\Gamma \vdash f : (\sigma_1,\,\ldots,\,\sigma_m) \rightarrow (\tau_1,\,\ldots,\,\tau_n)
		&\quad&
		\Gamma \vdash x_1 : \rho_1 \, \ldots \, \Gamma \vdash x_n : \rho_n\\
\end{array}
}{\Gamma \vdash f(x_1,\,\ldots,\,x_m) : (\tau_1,\,\ldots,\,\tau_n)}{app}
}
\end{array}$$
\caption{Typing rules for \ATF{} expressions}\label{exptypes}
\end{figure}

In \figref{statypes}, $CorrectStatement$ denotes all the statements in which typing rules are respected. 

\begin{figure}[htbp]
$$
\begin{array}{l}
\trule{
\begin{array}{l}
\sigma,\,\tau \in \G \cup \TUP
\\
\Gamma \vdash \alpha : \sigma 
\quad 
\Gamma \vdash \beta : \tau 
\quad 
\sigma \gel^* \tau
\end{array}
}{\Gamma \vdash \alpha = \beta \in CorrectStatement}{assignment}
\vspace{10pt}
\\
\trule{
\Gamma \vdash f(x_1,\,\ldots,\,x_m) : \rho
}{
\Gamma \vdash f(x_1,\,\ldots,\,x_m) \in CorrectStatement
}{app-statement}
\vspace{10pt}
\\
\trule{
\Gamma \vdash \alpha_1,\,\ldots,\,\alpha_n \in CorrectStatement
}{\Gamma \vdash \left\{ \alpha_1,\,\ldots,\,\alpha_n \right\} \in CorrectStatement}{block}
\end{array}
$$
\caption{Type checking rules for \ATF{} statements}\label{statypes}
\end{figure}

The only nontrivial constraint is expressed by the rule \textsc{assignment}: the type of the right-hand side must be a subtype of the type of the left-hand side. As the left-hand side may appear to be a tuple (as well as the right-hand side in case of function application), we use an extended subtyping relation $\gel^*$ which is the minimal relation such that $\gel \subset \gel^*$ and
$$(a_1,\cdots,a_n)\gel^*(b_1,\cdots,b_n)\mbox{ iff } (a_1 \gel b_1) \wedge \cdots \wedge (a_n \gel b_n)$$

\subsection{Type inference.}\label{TypeInference}

The previous subsection formalizes the type system of \ATF{} under the assumption that every attribute and every function which is used is also declared explicitly. As we illustrated above, such declarations may be redundant in some cases, and \ATF{} (after many programming languages) provides a local type inference mechanism to enable omission of some of them.  

Type inference works separately in each translation function. It represents all the statements as sequences of attribute assignments (function arguments are treated as ``assigned'' to input attributes) assigning unknown types to undeclared attributes. To reconstruct the declarations we use a modification of a conventional algorithm (see, for example, \cite{Pierce}) which creates constraints (subtyping inequalities) for unknown types and finds ground types which satisfy these constraints (see \secref{TypeInf} for a simplistic example). Our modifications to the classical algorithm are not significant enough to formally present the entire type inference process. Instead, we will only illustrate the behaviour of our algorithm in case of an ambiguity.

Let us assume that the implementation language has the following types: $\G = \{Object, Integer\}$, where $Integer \lel Object$. Consider the following function:
\begin{lstlisting}[language=Grammatic]
f(Integer x) --> (Object result) {
	before ... : t = x;
	// ...
	after  ... : result = t;
}
\end{lstlisting}
The local attribute \texttt{t} is not declared. Let $\tau$ be the unknown type for \texttt{t}. The type inference algorithm will construct the following set of constraints:
$$\left.\begin{array}{l}
Integer \lel \tau\\
\tau \gel Object\\
\end{array}\right\}$$
These constraints have at least two solutions: both $Object$ and $Integer$ might be assigned as a type for \texttt{t}. The very existence of a solution already means that the specification does not have type inconsistencies, but the back-end will need the exact type information to generate code, so we have to decide which type to choose. This may be important when inferring the types for external functions since they will be visible to the user (we provide details on the back-end below).

In such cases \ATF{} prefers lower bounds to upper bounds, which means that \texttt{t} will be assigned the type $Integer$. The general procedure is the following: find a minimal solution satisfying all lower bounds, if it also satisfies all upper bounds, choose it as a final solution. If there are several (incomparable) minimal solutions for the lower bounds which satisfies all the upper bounds, the algorithm can not decide between them and reports an error. If no solution for lower bounds satisfies all the upper bounds, the specification is inconsistent, and we again report an error. If no lower bounds are present, we choose a \emph{maximal} solution for the upper bounds. 

This procedure can be summarized as follows: we look for the type which is as close to the constraining ones as possible, preferring more concrete (smaller) types. This approach appears to be rather intuitive: it is very unlikely to infer a type which the developer does not expect.

If no constraints are present at all, this means that all the attributes connected to the one at hand are not declared. In this case we ask if the ground type system has a top type $Top_L$ (such as \texttt{java.lang.Object}), and if it has one, we choose it, otherwise we report an error.

\subsection{Revisiting the initial example.}

Now we are ready to explain how the example from \secref{Introduction} is handled by \ATF{}. In that example we had a rule for \texttt{factor} analogous to the one in \lstref{factorTF}, in which there was an error: a name of the \texttt{INT} token was used instead of its textual value (\texttt{INT\#}). Here is the modification of \lstref{factorTF} containing the same defect:
\begin{lstlisting}
factor : VAR | INT | '(' expr ')' ;          
	factor(Environment env) --> (Int result) { 
		after VAR : result = evaluate(env, VAR#);  
		after INT : result = INT;
		at expr   : result = expr(env);           
	}
\end{lstlisting}
Unlike \tool{ANTLR}, \ATF{} reports the following error when we try to generate code: \emph{``The local attribute INT might have not been initialized''}. What happened? \texttt{INT} appears on the right-hand side of an assignment. Thus, \ATF{} expects it to be an attribute. Since such an attribute is not declared, the type checker treats it as a local attribute and infers a type for it: \texttt{Int}. For the time being, no error has been found. After the type checking, the definitive assignment analysis is performed, and it finds that the local attribute \texttt{INT} is read but never assigned. This leads to the error which is reported.

Without generating the code and running a compiler, we have got an error message which points precisely to the place in the specification where the defect is situated. Following the logic of the example of \secref{Introduction}, we correct the specification:
\begin{lstlisting}
	after INT : result = INT#;
\end{lstlisting}
Now the type checker complains: \emph{``Incompatible types: String and Int''}. Again, we have got a precise error message without generating code and running a compiler. We correct the specification again:
\begin{lstlisting}
	after INT : result = strToInt(INT#);
\end{lstlisting}
This time all checks are passed successfully, the code is generated and will be compiled with no errors.

As can be seen, the development cycle now takes the form shown in \figref{cycles} (right side), which was our main goal. Now we proceed to a description of the tools \ATF{} provides to support many implementation languages.


\section{Declarative descriptions of type systems.}\label{DefaultExtension}

The type checking procedure described above is parameterized by a set of ground types $\G$ with distinguished types $\Str$ and $Top_L$, and a subtyping relation~$\lel$. To incorporate a type system of the implementation language into \ATF{}, one needs to substitute concrete implementations for these parameters. In general, this is done by adding front-end \emph{extensions} (plug-ins written in Java), which provide support for particular implementation languages. Developing such extensions requires some effort and may be undesirable in certain cases. For this reason \ATF{} provides a default extension which supports declarative descriptions of type systems of implementation languages. 

A type system description specifies a set of named types, a subtyping relation on these types, an optional top type and a string type. For example, the type system used above may be described as follows:
\begin{lstlisting}[language=Typesystem]
	typesystem Simple( 
		_,             // Name of the top type (nothing in this case)
		String         // Name of the string type
	) {
		// Type declarations
		type Int;
		type Environment;
		type Object;

		// Subtyping rules
		Environment <: Object;
		String      <: Object;
	}
\end{lstlisting}
This denotes a type system named \texttt{Simple}, with no top type (underscore is used to denote this, if there were a top type, its name would have been written) and with the string type called \texttt{String}. It declares three more types: \texttt{Int}, \texttt{Environment} and \texttt{Object}. The first two were used above, but the third one is added only for demonstration purposes, namely to introduce subtyping rules which state that \texttt{Environment} and \texttt{String} are subtypes of \texttt{Object}. Note that since \texttt{Object} is not the top type, \texttt{Int} is not its subtype. In a general case, subtyping rules stated in a type system description form an incomplete form of a subtyping relation. A final subtyping relation is obtained as a reflexive-transitive closure of it.

Type system descriptions such as the one shown above are sufficient to check \ATF{} specifications for type safety and infer types, but they are not sufficient to generate code unless the implementation language has types with exactly the same names as used in a description. The latter is unlikely because normally types are defined inside some kind of namespaces, such as Java packages, and can not be referred to by a simple name without import statements. Thus, we have to provide another description which \emph{instantiates} the types with their actual form in the implementation language. We call this a \emph{language description}:
\begin{lstlisting}[language=Typesystem]
	language Java for Simple {
		Int = 'int';
		Environment = 'java.util.Map<String, Integer>';
		String = 'String';
		Object = 'Object';
	}
\end{lstlisting}
The example shows a language description named \texttt{Java} which instantiates the type system \texttt{Simple} and says that \texttt{Int} is implemented as \texttt{int} in Java, and \texttt{Environment} is implemented as a map from strings to integer objects. We might not specify the instantiations for \texttt{String} and \texttt{Object} since they may be referred to just by their names.

Now a back-end can use the strings we have provided in generated code, and it will work correctly (if the back-end generates Java and not C or some other language). A back-end may need some extra information such as a package to put the generated code in and a name to give to the generated parser. These options are provided in a \emph{back-end profile}, such as
\begin{lstlisting}[language=Typesystem]
	backend 'org.grammatic.pg.backends.ANTLRJavaBackend' for Java {
		package = 'org.example.arithexp';
		parserName = 'ExpressionEvaluator';
	}
\end{lstlisting}

This is a profile for a back-end implemented by a Java class \texttt{ANTLRJavaBackend} which applies for the language description \texttt{Java} defined above. In the profile one simply writes name-value pairs which are processed by the back-end as options.

\begin{figure}[htbp]
		\includegraphics[width=\textwidth]{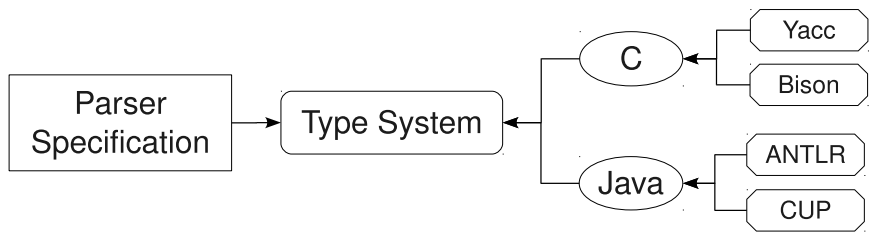}
		\caption{Three levels of type system descriptions}\label{typesystem}
\end{figure}

To summarize, the declarative descriptions of type systems in \ATF{} are organized into three levels (see \figref{typesystem}). 

This makes \ATF{} rather flexible when it comes to \emph{multi-targeted} parser specifications, form which parsers in many implementation languages must be generated. 
In such a case one describes an abstract \emph{type system} as shown above and provides many \emph{language descriptions} for it, so that each \emph{back-end profile} may use its own language. In some situations it is also convenient to have several back-end profiles for the same language, for example, when one needs to compare performance of different implementations or while migrating from one back-end to another.


\section{Language-specific front-end extensions.}\label{Java}

To provide tighter integration with a particular implementation language one can use a language-specific extension of the \ATF{} front-end instead of the default one described in the previous section. 
Technically, a front-end extension consists of a ground type syntax specification and implementations of Java interfaces which capture the semantical aspects of ground types: a subtyping relation, a set of predefined types and two distinguished types. Let us describe these parts in more details.
Examples below present an extension supporting Java types (including generics), which we developed in our prototype.

\subsection{Syntax of ground types.}

The core specification of the \ATF{} notation (see \lstref{short_notation}) has extension points: it uses but does not define two nonterminal symbols, \texttt{type} and \texttt{declaration} (written in bold font in the listing). The language generated by \texttt{type} is a syntactical form of~$\G$. Since the specification parser in \ATF{} is itself implemented using \ATF{}, \texttt{type} must be defined by grammar rules and translation functions. This is done in a separate specification file which is virtually ``appended'' to the generic specification when the whole system is built.
\begin{lstlisting}[caption={Extension points in \ATF{} notation},float=htbp,label=short_notation,
		emph={declarations,type},emphstyle=\bfseries,
%		emph={[2]grammarRule,NAME,STRING},emphstyle={[2]\it}
]
specification 
	: declarations? 
	  (externalFunctionSignature | (grammarRule translationFunction*))* ;
attributeDeclaration 
	: type NAME ;
\end{lstlisting}

A syntactic function for \texttt{type} must return an instance of \texttt{java.lang.Object} (\ATF{} is implemented in Java), in other words, a type may be represented by an arbitrary object. The context-free rules for types in Java~5 are given in \lstref{java_types}. For the sake of brevity we do not include the corresponding translation functions.

\begin{lstlisting}[caption=Ground type syntax for Java 5,float=htbp,label=java_types]
type
	: IDENTIFIER typeArguments? ('.' IDENTIFIER typeArguments?)* ('[' ']')*
	: basicType ;
typeArguments
	: '<' typeArgument (',' typeArgument)* '>' ;
typeArgument
	: type
	: '?' (('extends' | 'super') type)? ;
basicType
	: 'byte' | 'short' | 'char' | 'int' | 'long' | 'float' | 'double' 
	| 'boolean';
\end{lstlisting}

With the rules from \lstref{java_types} used for defining the syntax of ground types, a signature of the \texttt{evaluate} function may be the following:
\begin{lstlisting}
evaluate
	(java.utli.Map<java.lang.String, java.lang.Integer> environment) 
	--> 
	(java.lang.Integer result);
\end{lstlisting}
As can be seen, fully qualified class and interface names and generics can be used as types for attributes. 

\subsection{Declarations.}

The names in the example above are quite long which is inconvenient. In Java this problem is solved with the help of imports. \ATF{} does not know about the structure of Java types and can not support imports itself. Instead, it provides a generic mechanism for adding arbitrary \emph{declarations} which are specific for an implementation language. The syntax for declarations is defined by the \texttt{declarations} nonterminal. A translation function for it does not take or accept any attributes: it is supposed to collect the information about the declarations and store it internally to be available when somebody else (e.g., the translation function for \texttt{types}) needs it.

To support imports we can define \texttt{declarations} as follows:
\begin{lstlisting}
declarations
	: importDeclaration* ;
importDeclaration
	: 'import' IDENTIFIER  ('.' INDENTIFIER)* ('.' '*')? ;
\end{lstlisting}
The corresponding translation functions will collect the information about imported types and provide it to the translation function for \texttt{types}. Now we can use Java imports in \ATF{} specifications, for example
\begin{lstlisting}[language=Java]
import java.util.Map;

evaluate(Map<String, Integer> environment) --> (Integer result);
\end{lstlisting}
 
The complete syntax of declarations which we use for Java is given in \lstref{java_decl}. In addition to imports, it supports \emph{options} which are used by the back-end and specify auxiliary information (this corresponds to back-end profiles of the default extension).
\begin{lstlisting}[caption=Declaration syntax for Java 5,label=java_decl,float]
declarations
	: options? importDeclaration* ;
options
	: '#javaoptions' '{' option+ '}' ;
option
	: NAME '=' STRING ';' ;
importDeclaration
	: 'import' IDENTIFIER  ('.' INDENTIFIER)* ('.' '*')? ;
\end{lstlisting}

\subsection{Semantics of ground types.}

\begin{lstlisting}[language=Java,caption=Java interfaces describing semantics of ground types,label=java_intf,float]
public interface ISubtypingRelation<T> {
	boolean isSubtypeOf(T type, T supertype);
}

public interface ITypeSystem<T> {
	ISubtypingRelation<T> getSubtypingRelation();
	Set<T> getPredefinedTypes();
	T getTopType();
	T getStringType();
}
\end{lstlisting}

Semantics of ground types is provided by Java classes that must implement the interfaces shown in \lstref{java_intf}.

A subtyping relation is represented by a class which implements \texttt{ISubtypingRelation<T>} interface, where the type parameter must be substituted by a class which is used by the extension to internally represent types, and the \texttt{isSubtypeOf} method returns true if and only if the following condition holds:
$$\mathtt{type} \lel \mathtt{supertype}$$
For example, our implementation represents Java types using the \texttt{EGenericType} abstraction from Eclipse Modeling Framework \cite{EMF}. In this case the subtyping relation class is declared as follows:
\begin{lstlisting}[language=Java]
public class JavaSubtypingRelation 
		 	 implements ISubtypingRelation<EGenericType>
\end{lstlisting}

The other interface, \texttt{ITypeSystem} has methods which return a subtyping relation represented as discussed above, a set of predefined types, a top type (or \texttt{null} if no top type exists in the ground type system) and a type for character strings. A back-end must use the \texttt{toString()} method of type objects to obtain their textual representation.

\section{Back-end.}\label{Backend}

By now we have presented an extensible front-end which can detect typing errors in specifications. Here we will explain why detecting these errors is sufficient for the back-end to be able to generate error-free code. The techniques described below apply to virtually any language, and we believe that whenever the front-end can be extended to support a particular implementation language, a corresponding back-end can be developed \footnote{Currently a back-end must be programmed manually in Java. This is because the issues described in this sections are rather peculiar for each implementation language and for the moment we do not see an efficient way to abstract them into a reusable framework. Such a framework is an interesting direction for the future research.}.

Generating error-free code from a grammar with no semantic actions is relatively easy. The problems arise when we need to incorporate hand-written code fragments into the generated program. \ATF{} front-end guarantees that the actions do not contain errors themselves and thus the errors may be caused only by conflicts between hand-written and generated code. For example, semantic actions may introduce names which are already used in the generated code or require particular imports. 

The peculiar property of this sort of errors is that the back-end can always detect and prevent them while reading the internal representation of the specification. This is because the back-end has total control over the generated code. For example, to avoid name clashes, it is sufficient to rename variables, which the back-end can do.

In the case when the back-end generates \tool{ANTLR} specifications with semantic actions in Java, we have to prevent the following types of errors:
\begin{itemize}
	\item A name used in the specification is a Java or ANTLR keyword.
	\item A name is used internally by ANTLR.
	\item A type requires some classes or interfaces to be imported.
\end{itemize}
Other types of errors are either prevented by the analysis performed by the front-end (e.g., usage of uninitialized variables) or do not appear because of the particular structure of the code generated by \tool{ANTLR}.

Naming problems are easy to prevent since it is sufficient to use a fresh name which can be obtained by adding numbers to original names (e.g., \texttt{result1}). The code remains readable enough and no errors appear.  
The import problem is also fixed straightforwardly: we can always import all the classes ever mentioned in the specification or use fully qualified names if some short names clash.

Correctness of external function signatures is guaranteed by the following design of generated parsers. Along with an \tool{ANTLR} specification, \ATF{} generates a Java interface which has a method for each external function used in the specification. This interface must be supported in order to implement the functions, which makes Java compiler to check if the functions are declared properly in the implementation. This interface being read and implemented by a human makes us select the most appropriate types during the type inference process (see \secref{TypeInference}).

To summarize, we have demonstrated how a back-end can prevent all the errors which are imposed by the structure of the generated code (and thus not present in the specification, and not checked by a front-end). These techniques vary over implementation languages, but the essence stays unchanged: we can always generate code in such a way that these errors do not appear.
Thus, we have reached our goal: as soon as a specification is successfully type-checked, the generated code is error-free.


\section{Related work.}\label{RelatedWork}

We are not aware of any parser generators which support multiple implementation languages and have type checking in their front-ends.

The most popular tools supporting multiple implementation languages are \tool{ANTLR} \cite{ANTLR} and \tool{Coco/R} \cite{Coco/R}, and they do not perform any type checking in the front-ends. In most cases these tools can not generate code in different specification languages from the same specification (the specifications, thus, are not \emph{multi-targeted} in these systems) , because the embedded actions are written in a particular implementation language and will not compile in another one. In \tool{SableCC} \cite{SableCC} specifications are multi-targeted, which is achieved by having no semantic actions: a developer has to manually process parse trees using visitors. In contrast, \ATF{} supports both multi-targeted specifications (using type system descriptions) and semantic actions.

The following attribute grammar systems are capable of reducing the development cycle to the one shown in \figref{cycles} (right side): \tool{Eli} \cite{Eli} automatically tracks compiler errors back to the specification. This approach is tied up not only to a specific implementation language (\tool{Eli} uses C), but also to a specific implementation of its compiler, since the format of error messages usually varies from one compiler to another.
The team behind \tool{JastAdd} \cite{JastAdd} system plans to integrate their own implementation of a Java compiler into it, to check semantic actions. This approach is also tied up to a specific implementation language. None of these systems provide appropriate means by which they could be extended for other implementation languages. \ATF{} allows for this via front-end extensions and separated back ends.


\section{Conclusion.}\label{Conclusion}

This paper addresses the problem of type checking in front-ends of parser generators supporting multiple implementation languages. The main goal is to prevent typing errors in the generated code to avoid the need of manually tracing such errors back to their causes in the specification . 

We have demonstrated that type checking of the specifications, which we implemented in a prototype tool \ATF{}, helps to reduce the development cycle compared to the one imposed by the tools currently available (see \figref{cycles}). Our approach is designed to be extensible for use with multiple implementation languages.

The principle contributions of this paper are the following:
\begin{itemize}
	\item A \ATF{} specification language supporting 
		\begin{itemize}
			\item semantic actions, but having no problem of tangling between grammar rules and action code;
			\item extensions to support type systems of implementation languages.
		\end{itemize}
	\item A type checking procedure for this language, supporting local type inference, compatible with the extensions.
	\item A generic extension for declarative definitions of abstract type systems, their syntactical realizations for particular language and configuration profiles for different back-ends, which in complex enable multi-targeted specifications.
	\item Another extension providing tight integration with Java.
\end{itemize}

We have also reported on a prototype back-end for generating \tool{ANTLR}/Java, which, we believe, never produces erroneous code from successfully type-checked specifications.

One possible direction to continue this work is to investigate a possibility of declaratively specifying back-ends for particular implementation languages to obtain a complete generator from declarative specifications. Another direction will be to integrate grammar inspections (such as heuristic ambiguity tests) into the static checking procedure.


\section*{Acknowledgements.}
This work was partly done while the author was a visiting PhD student at University of Tartu, under a scholarship from European Regional Development Funds through Archimedes Foundation.


\begin{backmatter}
\begin{biography}
\author{A. Breslav} received his Masters degree in Computer Science from St. Petersburg State University of IT, Mechanics and Optics in 2007. He is currently a Ph.D. student at the Mathematics department of the same university and a visiting Pd.D student at the Institute of Computer Science of University of Tartu. His research interests include the problems of declarative language descriptions, type systems and domain-specific languages.
\end{biography}

\end{backmatter}
\end{document}